\documentclass[12pt]{iopart}

\usepackage[dvips]{graphicx}
\def\la{\langle}
\def\ra{\rangle}
\def\s{\sigma}

\begin{document}

\title{Continuum scaling in expansions effective at a large lattice spacing}

\author{H Yamada}

\address{Division of Mathematics, Chiba Institute of Technology, Shibazono 2-1-1, Narashino, Chiba 275-0023, Japan}
\ead{yamada.hirofumi@it-chiba.ac.jp}
\begin{abstract}
A new class of truncation schemes of delta expansion on the lattice is studied.  We show that the order of expansion in $\delta$ which is introduced as the dilation parameter can be taken large enough and the result gives rise to the Borel transformation with respect to the relevant variable in the lattice models.   The explicit simulation of the continuum scaling from the expansion effective at large spacings is investigated in anharmonic oscillators, $d=2$ non-linear $\sigma$ model at large ${\cal N}$ and Gross-Neveu model with Wilson fermions. 
\end{abstract}

\maketitle

\section{Introduction}
\label{intro}
From long time ago, many papers have appeared on the nonperturbative computational technique of quantum systems based on the idea of interpolating actions \cite{vast}.   The technique is called under various names such as variational purterbation, Gaussian approximation,  linear delta expansion, interpolation method and so on.  All of these methods exploit artificially incorporated parameter $\delta$ which interpolates non-trivial action of interest (at $\delta=1$) and some solvable one (at $\delta=0$).  When $\delta$ is set to unity, the total action becomes independent of the solvable one and the independence is often utilized to obtain non-trivial results by the use of principle of minimum sensitivity \cite{pms}.

In some cases of those approaches, the introduced parameter $\delta$ can be viewed as the dilation parameter for one of parameters originally included.  On the point of view of dilation, one can use the method in new ways.  For example, in lattice field theories, we can start with the strong coupling expansion which is often valid at a large lattice spacing $a$ and then perform the expansion in the dilation parameter $\delta$ after the shift of $a\to a(1-\delta)$.   This technique, simply called delta expansion, was proposed and applied to some lattice models and it was shown that the continuum scaling emerges in the strong coupling series \cite{yam,hhy}.  The non-perturbative mass gap in the continuum limit was then evaluated to good accuracy.   To be self-contained, we first describe the method below.

Let a quantity of interest $\Omega$ has an expansion:
\begin{equation}
\Omega=\sum_{k=1}\frac{\omega_{k}}{M^{k}}.
\label{largem}
\end{equation}
Here, $M$ denotes the mass square in momentum space rescaled to be dimensionless.  The continuum scaling of $\Omega$ is given by the behavior as $M\to 0$.  
  To access the small $M$ behavior of $\Omega$, we dilate $\Omega(M)$ around the continuum limit by shifting $M$ as $M\to M(1-\delta)$ where $0\le \delta\le 1$.   Suppose then
\begin{equation}
\Omega(M)\sim A M^{-\alpha},\quad (\alpha>0),
\label{scaling}
\end{equation}
in the scaling region.  Dilation leads that $\Omega(M(1-\delta))\sim A M^{-\alpha}(1-\delta)^{-\alpha}$ and as $\delta\to 1$ dilated $\Omega$ diverges for any fixed $M>0$ as in the manner that $\Omega(M)$ diverges as $M\to 0$.  Since the original scaling behavior is thus transferred to the behavior of dilated function in the $\delta\to 1$ limit, there is a possibility that the scaling behavior could emerge in dilated large $M$ expansion of $\Omega$, even if we have only finite series to order $M^{-K}$.  To put this idea into practice, a critical step is to expand $\Omega(M(1-\delta))$ in dilation parameter $\delta$ both at large and small $M$ in an appropriate manner.  Now, $K$th order $\Omega(M(1-\delta))$ at large $M$ reads
$\Omega(M(1-\delta))=\sum_{k=1}^{K}\frac{b_{k}}{M^k(1-\delta)^k}$.  In \cite{yam,hhy}, the expansion in $\delta$ is truncated along with the conventional rule of partial sum,
\begin{equation}
\Omega((1-\delta)M)\sim \sum_{1\leq i+j\leq K} \omega_{ij}M^{-i}\delta^{j}.
\label{triangle}
\end{equation}
Here $\omega_{1j}=\omega_{1}, \omega_{2j}=(j+1) \omega_{2}, \omega_{3j}=\frac{(j+2)(j+1)}{2}\omega_{3}, \cdots$.  
The order of term $M^{-i}\delta^j$ is assigned as $i+j$ and it is included as long as $i+j\le K$.  Note that maximum order of $\delta$ is $K-1$.   
On the other hand, at small $M$, we need only the leading term (\ref{scaling}) giving asymptotic scaling.  We expand it in $\delta$ to $\delta^{K-1}$ to keep accordance with the highest power of $\delta$ of (\ref{triangle}).  The result then reads
\begin{equation}
\Omega((1-\delta)M)\sim A M^{-\alpha}\sum_{i=0}^{K-1}\frac{\alpha(\alpha+1)\cdots(\alpha+i-1)}{i!}\delta^{i}.
\label{smallm}
\end{equation}
Setting $\delta=1$ which means the dilation to infinite amount, the truncated series of $\Omega$ at $M\gg 1$ and $M\ll 1$ were compared with each other and (\ref{triangle}) was found to exhibit the correct "scaling behavior" (\ref{smallm}) at $\delta=1$.   We emphasize that, due to dilation, the scaling is observable in a wide region of shifted $M$ of order $M\sim O(1)$ or even larger. 

As explicit in (\ref{triangle}) the series at large $M$ has two expansion parameters, $M^{-1}$ and $\delta$.  If both parameters are small enough, the truncation of (\ref{triangle}) would be the most reasonable.  However, $\delta$ is always set to $1$ eventually.  Then, terms of different $j$ for fixed $i$, $b_{ij}(M^{-1})^i\delta^j$, mix among themselves.  Hence at $\delta=1$ the order assignment loses the basis.  This motivates us to seek for and consider other truncation schemes admitting analytic control.   Within those schemes, we like to examine convergence properties as the order of expansion increases.  In addition we will point out that when $M$ and the order $N$ of the expansion in $\delta$ is large enough with $M/N$ kept fixed, large $M$ series approaches to the Borel transform of the original series (\ref{largem}) \cite{yam2}.  Thus, by exploring various possible schemes, we can obtain computational flexibility toward models complex and not fully understood yet.  As a demonstration of important role of Borel transform limit, we reinvestigate the continuum limit of the Gross-Neveu model with Wilson fermion.

This paper is organized as follows:   In the next section, we introduce a class of new truncation schemes which we call "square" schemes.  Then we formally show that those schemes lead to Borel transform in a certain limit.  In section 3, delta expansion in square schemes is examined in detail by applying the schemes to a simple model, the ${\cal N}$ component anharmonic oscillator in the large ${\cal N}$ limit.  Then, we apply Borel transform to study the scaling properties and evaluate the mass gap.  In section 4, continuum scaling of three models, ordinary anharmonic oscillator which corresponds to ${\cal N}=1$ case, two dimensional (2d) non-linear $\sigma$ and Gross-Neveu model with Wilson fermion are studied at large ${\cal N}$ in the Borel transform approach.  In the non-linear $\sigma$ model, to improve the accuracy of approximating the continuum limit, Symanzik's improved action \cite{sym} will be discussed to the first order.  In Gross-Neveu model, we point out that the conventional truncation scheme fails to capture the continuum scaling but the new one works good.  The results of this work is summarized in the last section.  In the study of anharmonic oscillators, we will confine ourselves with the pure anharmonic case where the harmonic mass term is absent from the action.

\section{Delta expansion in square truncation and its certain limit leading to Borel transform}
\label{sec:1}
In this section we define square schemes and study the limit where the order of expansion in $\delta$ is taken to large enough.

To define a square scheme, let $\Omega(M(1-\delta))$ be first expanded in $\delta$ to $\delta^{N}$ such that $\Omega(M(1-\delta))=\Omega_{0}(M)+\Omega_{1}(M)\delta+\Omega_{2}(M)\delta^2+\cdots+\Omega_{N}(M)\delta^{N}$ ($\Omega_{0}(M)=\Omega(M)$).  At $M\ll 1$, expanding $\Omega_{k}$ in $M$ and collecting leading terms for each $k$, the result (\ref{smallm}) with the replacement $K-1\to N$ is derived.  Then, we find in the $\delta\to 1$ limit
\begin{equation}
\Omega(M(1-\delta))\to A M^{-\alpha}Z_{N}(\alpha)
\label{scalealpha}
\end{equation}
where
\begin{equation}
Z_{N}(\alpha)=\frac{\Gamma(N+\alpha)}{(N-1)!\Gamma(\alpha+1)}.
\end{equation}
At $M\gg 1$, we expand $\Omega_{k}$ in powers of $M^{-1}$ to $(M^{-1})^{K}$.  Then at $M\gg 1$, the partial sum of dilated $\Omega$ is formally written as
\begin{equation}
\Omega((1-\delta)M)\sim \sum _{1\leq i\leq K, 1\leq j\leq N}\omega_{ij}M^{-i}\delta^j.
\label{square}
\end{equation}
The order of the truncated double series (\ref{square}) are labeled by $K$ and $N$.  Finally we set $\delta=1$ in (\ref{square}).   To obtain (\ref{square}) at $\delta=1$ from the original $1/M$ series (\ref{largem}), we suffice to expand dilated term $M^{-k}(1-\delta)^{-k}$ 
to the order $N$ such that $M^{-k}(1+k\delta+k(k+1)/2!\delta^2+\cdots+k(k+1)\cdots(k+N)/N!\delta^{N})$.  Then, setting $\delta=1$, we obtain the simple transformation rule,
\begin{equation}
M^{-k}\to M^{-k}\Big(
\begin{array}{c}
N+k \\
k
\end{array}
\Big).
\end{equation}
This gives the dilated $\Omega$ at $\delta=1$ to orders $(K, N)$ as
\begin{equation}
\sum_{k=1}^{K}\frac{b_{k}}{M^k}\quad \to \quad  \sum_{k=1}^{K}\Big(
\begin{array}{c}
N+k \\
k
\end{array}
\Big)\frac{b_{k}}{M^k}:=D\Big[\sum_{k=1}^{K}\frac{b_{k}}{M^k}\Big].
\label{sdelta}
\end{equation}
 
We always deal with the truncated series and there appears no divergence connected to the $M\to 0$ limit even when $\delta$ is set to unity.   Though $\Omega(0\times M)$ is $M$ independent, the right hand side of (\ref{scalealpha}) and (\ref{sdelta}) have $M$ dependence.  As the order of expansion increases,  the residual $M$ dependence should become weaker. 

 In 2d non-linear $\sigma$ model,  we consider the scaling of bare coupling ($\Omega=\frac{1}{g}$).  Then $\alpha=0$ in (\ref{scalealpha}) and the scaling behavior is a logarithmic one, $\Omega(M)\sim A\log M+B$.  Since $\log M$ turns to $\log M(1-\delta)$ under the dilation and is expanded as $\log M-\sum_{k=1}^{N}\frac{\delta^{k}}{k}$, we have
\begin{equation}
\log M\to \log M-\sum_{k=1}^{N}\frac{1}{k},
\end{equation}
where $\delta$ has set to $1$.  
Hence, we have the following modification of the continuum logarithmic scaling,
\begin{equation}
A \log M +B \to A\Big(\log M-\sum_{k=1}^{N}\frac{1}{k}\Big)+B.
\label{smalllog}
\end{equation}
By calculating (\ref{sdelta}) and comparing it with (\ref{scalealpha}) or (\ref{smalllog}), the informations on $A$ and $B$ can be extracted.

Now, we consider the limit that the order $N$ is taken large enough.  In the limit, the factor $\Big(
\begin{array}{c}
N+k \\
k
\end{array}
\Big)$ behaves as
\begin{equation}
\Big(
\begin{array}{c}
N+k \\
k
\end{array}
\Big)\sim \frac{N^k}{k!}.
\end{equation}
Hence $M^{-k}$ in original expansion transforms to $\frac{1}{k!}(M/N)^{-k}$.  We then obtain
\begin{equation}
D\Big[\sum_{k=1}^{K}\frac{b_{k}}{M^k}\Big]\to  \sum_{k=0}^{K}\frac{1}{k!}\frac{b_{k}}{(M/N)^k}.
\label{borel}
\end{equation}
If $M$ is $O(1)$, all terms in the right-hand-side (RHS) of (\ref{borel}) diverge.  However, 
if $M$ is taken large as the same order of $N$, the series in the RHS of (\ref{borel}) becomes sensible.  Thus, it is tempting to consider the following correlated limit,
\begin{equation}
M, N\to \infty\quad {\rm with}\,\, \frac{M}{N}\,\,{\rm fixed}.
\label{limit}
\end{equation}
By defining a new variable $\hat M$ by
\begin{equation}
\hat M:=\frac{M}{N},
\end{equation}
we summarize the result of square scheme to large orders as
\begin{equation}
\sum_{k=1}^{K}\frac{b_{k}}{M^k}\quad\to\quad\sum_{k=1}^{K}\frac{1}{k!}\frac{b_{k}}{\hat M^k}=B\Big[\sum_{k=1}^{K}\frac{b_{k}}{M^k}\Big].
\label{largeborel}
\end{equation}
It is obvious that the obtained result is nothing but that of Borel transform with respect to $M$.  On the other hand, at small $\hat M$, we find
\begin{equation}
\Omega(0\times M)\to A\times \hat M^{-\alpha}\frac{1}{\Gamma(1+\alpha)}
\label{smallborel}
\end{equation}
in the limit (\ref{limit}).  
For the case of the logarithmic scaling, we find from $\sum_{k=1}^{N}\frac{1}{k}\to \log N+\gamma_{E}$ ($\gamma_{E}$ stands for the Euler constant) that
\begin{equation}
\log M\to \log\hat M-\gamma_{E}.
\end{equation}
Hence,
\begin{equation}
\Omega(0\times M)\to A(\log\hat M-\gamma_{E})+B.
\label{borellog}
\end{equation}

The above argument leading to Borel transform is a formal one and needs detailed explanation.  Taking the ${\cal N}$ component anharmonic oscillator as an example, we like to demonstrate that the limit (\ref{limit}) is legitimate and sufficient for our purpose (see the next section).  

Note that $N$ is completely combined with the mass parameter $M$.  The effective variable is $\hat M=M/N$ and we need not specify the value of $N$, which plays the role of the regulator of the delta expansion.  By the comparison of (\ref{largeborel}) to (\ref{smallborel}) or (\ref{borellog}), we can obtain the informations of the exponent $\alpha$, constants $A$ and $B$, if (\ref{largeborel}) exhibits the scaling behavior at a region of small $\hat M$.

\section{Application to the ${\cal N}$ component anharmonic oscillator}
As the first application of square delta expansion and Borel transform (BT) to concrete models, we address to ${\cal N}$ component anharmonic oscillator in the large ${\cal N}$ limit, since the model is solvable within self-consistent approach.    The lattice spacing is $a$ and ${\cal N}$ component field $\vec{\phi}_{n}=(\phi_{n}^{1}, \phi_{n}^2,\cdots,\phi_{n}^{\cal N})$ denotes the dynamical degree on a cite $n$ $(n=0,\pm 1\pm 2,\cdots, \pm \bar{L})$.  The massless action is then given by
\begin{equation}
S=\sum_{n=-\bar L}^{\bar L}a\bigg[\frac{1}{2}\Big(\frac{\vec{\phi}_{n+1}-\vec{\phi}_{n}}{a}\Big)^2+\frac{\lambda}{4{\cal N}}
\{\vec{\phi}_{n}^2\}^2\bigg].
\end{equation}
The action can be rewritten by rescaling fields from $\vec{\phi}$ to $\vec{\varphi}=(a\lambda/4)^{1/4}\vec{\phi}$, giving
\begin{equation}
S=\sum_{n=-\bar L}^{\bar L}(\beta \vec\varphi_{n}^2+\frac{1}{\cal N}\{\vec{\varphi}_{n}^2\}^2)-\beta\sum_{n=-\bar L}^{\bar L}\vec\varphi_{n}\cdot \vec\varphi_{n+1},
\label{anhop}
\end{equation}
where
\begin{equation}
\beta=\bigg(\frac{4}{\lambda a^3}\bigg)^{1/2}.
\end{equation}
The second term represents the hopping term by which nearest neighbour fields tend to align {\footnote{The expansion parameter $\beta$ in our hopping expansion is twice of $\kappa$ which is used as the conventional hopping expansion.}}.  

Now, the self-consistent method gives
\begin{equation}
m^2=\lambda\int_{-\pi/a}^{\pi/a}\frac{dp}{2\pi}\frac{1}{\frac{2}{a^2}(1-\cos pa)+m^2},
\end{equation}
where $m$ stands for the dynamical mass at spacing $a$.  Changing the integration variable from $p$ to $\theta$ by $p=\theta/a$ and defining the dimensionless mass variable by
\begin{equation}
M:=(ma)^2, 
\label{mass}
\end{equation}
we find
\begin{equation}
\beta=\left[\frac{4}{M}\int_{-\pi}^{\pi}\frac{d\theta}{2\pi}\frac{1}{2(1-\cos \theta)+M}\right]^{1/2}.
\label{self}
\end{equation}
It is easy to obtain exact $\beta$ as a function of $M$ by calculating the integration explicitly but it is not our aim.  The starting point of our argument must be of generally available.  It is a series expansion of $\beta$ in powers of $1/M$ which is obtainable from (\ref{anhop}).  However, in the present case, the result can be obtained easier from (\ref{self}):
\begin{equation}
\beta=\frac{2}{M}-\frac{2}{M^2}+\frac{5}{M^3}-\frac{15}{M^4}+\frac{195}{4M^5}-\cdots=\sum_{k=1}^{\infty}\frac{b_{k}}{M^k}.
\label{largem2}
\end{equation}
We emphasize that the self consistent method is used just to produce the large $M$ expansion in an efficient way.  
In the next subsection, using only the above series, we can address to the scaling behavior,
\begin{equation}
\beta\sim \sqrt{2}M^{-3/4},
\label{scaling2}
\end{equation}
which is derived also from (\ref{self}).

\subsection{Delta expansion in a square truncation}
In this subsection, we use the method of delta expansion in a square truncation scheme to capture the scaling behavior (\ref{scaling2}) in the large $M$ series.   We proceed by correlating the orders of $1/M$ and $\delta$ expansions by
\begin{equation}
K=N.
\label{square1}
\end{equation}
Then, to the full order $K$, we have
\begin{equation}
D[\beta]=\sum_{k=1}^{K}\Big(
\begin{array}{c}
K+k \\
k
\end{array}
\Big)\frac{b_{k}}{M^k}.
\label{square1large}
\end{equation}
Figure 1 shows the plots of  $\beta$ (\ref{largem2}) and $D[\beta]$ (\ref{square1large}) from 2nd to 8th orders.  As explicit from Figure 1(2), $D[\beta]$ shows rough scaling at rather large $M$. 
\begin{figure}[h]
\begin{center}
\includegraphics[scale=0.7]{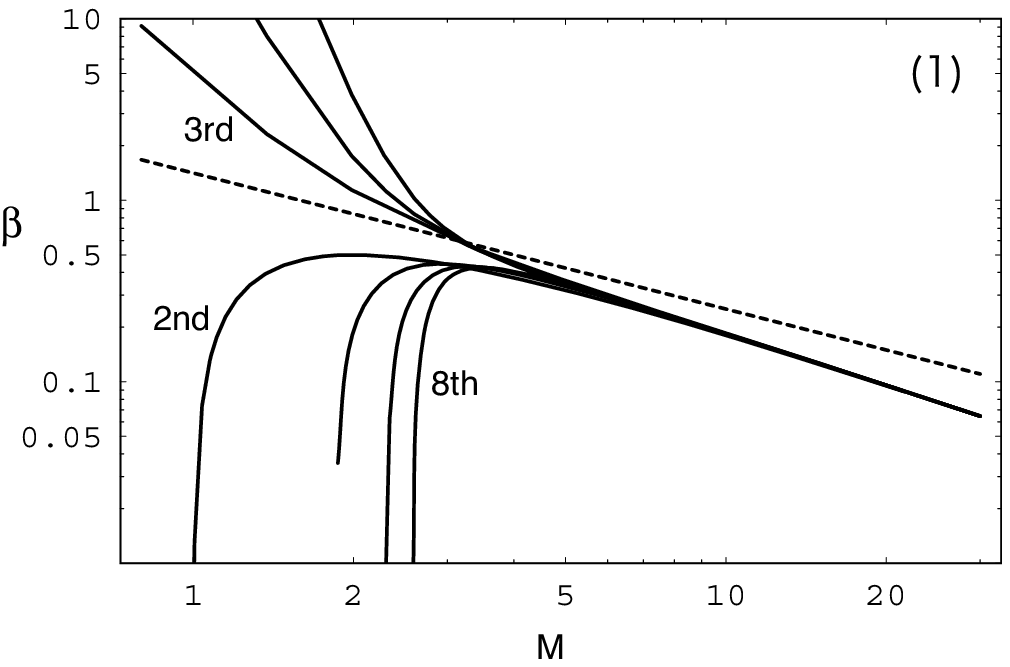}
\includegraphics[scale=0.7]{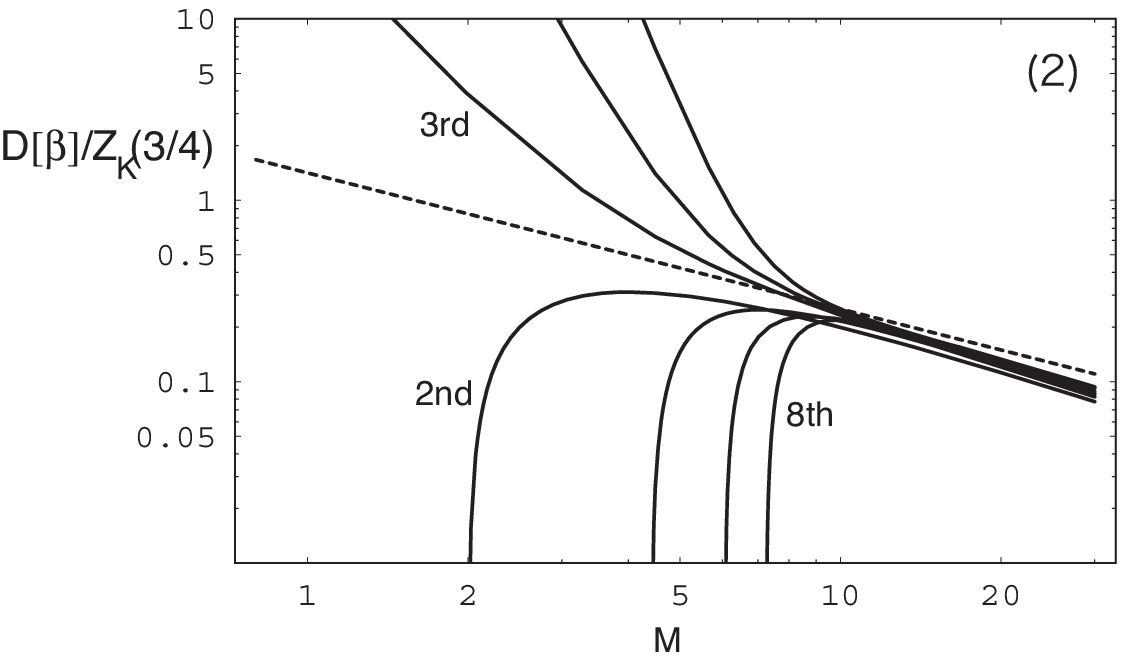}
\end{center}
\caption{(1) Plots of $\beta$ from 2nd to 8th orders.     (2) Plots of $D[\beta]/Z_{K}(3/4)$ from 2nd to 8th orders.  In both figures, each dotted line represents the asymptotic scaling $\beta\sim \sqrt{2} M^{-3/4}$.}
\end{figure} 
 
To access the continuum scaling, it is more convenient to deal with the logarithm of $\beta$.   This is because when the small $M$ behavior of $\beta$ is such as $\beta=AM^{-3/4}(1+c_{1}M+c_{2}M^2+\cdots)$, the delta expansion leaves all corrections to the asymptotic scaling to any finite orders in $\delta$.  However, for $\log\beta=\log A-3/4\log M+c_{1}M+O(M^2)$, the low order corrections, $M^1, M^2,\cdots$, are dilated and expanded as $M(1-\delta), M^2(1-2\delta+\delta^2),\cdots$, and they disappear by setting $\delta=1$ at several orders.  Now $\log \beta$ at large $M$ becomes
\begin{equation}
\fl \log\beta=\log \frac{2}{M}-\frac{1}{M}+\frac{2}{M^2}-\frac{16}{3M^3}+\frac{16}{M^4}-\frac{256}{5M^5}+\cdots=\log\frac{2}{M}+\sum_{n=1}\frac{b'_{n}}{M^n},
\label{30}
\end{equation}   
and
\begin{eqnarray}
\fl D[\log\beta]=\log \frac{2}{M}+\sum_{k=1}^{K}\frac{1}{k}-\Big(
\begin{array}{c}
K+1 \\
1
\end{array}
\Big)\frac{1}{M}+\Big(
\begin{array}{c}
K+2 \\
2
\end{array}
\Big)\frac{2}{M^2}-\Big(
\begin{array}{c}
K+3 \\
3
\end{array}
\Big)\frac{16}{3M^3}+\cdots\nonumber\\
=\log\frac{2}{M}+\sum_{k=1}^{K}\frac{1}{k}+\sum_{n=1}\Big(
\begin{array}{c}
K+n \\
n
\end{array}
\Big)\frac{b'_{n}}{M^n}.
\end{eqnarray}
It should be reminded that if $\beta$ is of order $K$, then the corresponding $\log\beta$ becomes of order $K-1$.  

At small $M$, $\log\beta\sim \log\sqrt{2}-\frac{3}{4}\log M$ and $D[\log\beta]\sim  \log\sqrt{2}-\frac{3}{4}(\log M-\sum_{k=1}^{K}\frac{1}{k})$.  
The behavior of $D[\log\beta]$ is plotted in Figure 2.  
\begin{figure}[h]
\begin{center}
\includegraphics[scale=0.7]{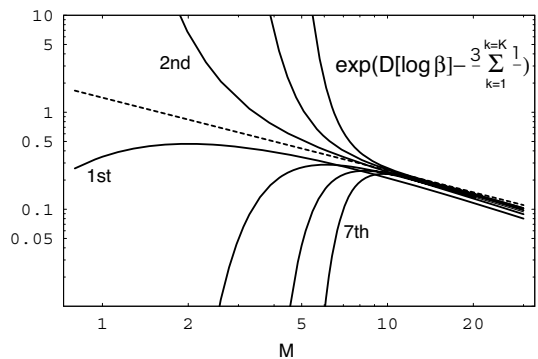}
\end{center}
\caption{Plots of $\exp(D[\log \beta]-\frac{3}{4}\sum_{k=1}^{K}\frac{1}{k})$ from 1st to 7th orders.  The dotted line represents the asymptotic scaling.}
\end{figure} 
It is clearly shown that we can observe the scaling behavior  in $1/M$ series more explicitly.  

To confirm the value of the exponent $\alpha$, we deal with $\frac{\partial \log\beta}{\partial\log M}$.  This function behaves at small $M$ as
\begin{equation}
\frac{\partial \log\beta}{\partial\log M}=-\alpha+\cdots
\end{equation}
where $\cdots$ represents the terms which vanish in the $M\to 0$ limit.  At large $M$ it behaves as follows:
\begin{equation}
\frac{\partial \log\beta}{\partial\log M}=-1+\frac{1}{M}-\frac{4}{M^2}+\cdots.
\end{equation}

If we introduce $\delta$ to dilate the region of $M$ by $M\to M(1-\delta)$ and take $\delta\to 1$ limit, $D[\frac{\partial \log\beta}{\partial\log M}]$ approaches to the uniform function taking the value $-\alpha$.  We note, however, when the dilated function is obtained as an approximant, it would have weak $M$ dependence and a plateau suggesting $-\alpha$.  By performing a square delta expansion on the large $M$ series of $\frac{\partial \log\beta}{\partial\log M}$, we have obtained the behavior shown in Figure 3.  At several orders, the truncated large $M$ series exhibits almost uniform behavior at large $M$, and the values at the plateaus are close to the exact value of $-\alpha=-3/4$.  Thus, the square delta expansion indeed shows correct value of $\alpha$. 

\begin{figure}[h]
\begin{center}
\includegraphics[scale=0.7]{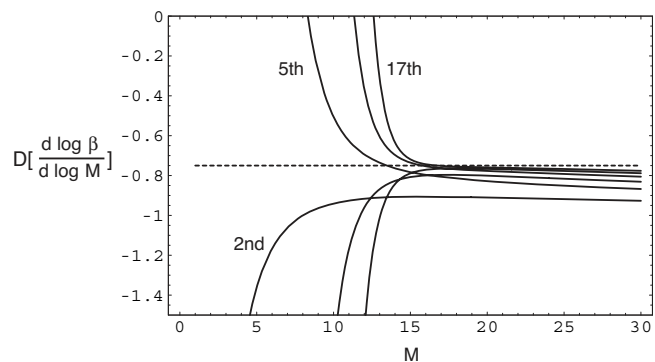}
\end{center}
\caption{Plots of $D[\frac{\partial \log \beta}{\partial \log M}]$ at $2$nd, $5$th, $8$th, $11$th, $14$th and $17$th orders.  The dotted line represents $-\alpha=-\frac{3}{4}$.}
\end{figure} 

The amplitude $\sqrt{2}$ in (\ref{scaling2}) can be evaluated as well.   From (\ref{scaling2}), $\log\beta+\frac{3}{4}\log M=Q(M)$ behaves in the scaling region,
\begin{equation}
Q(M)=\log\sqrt{2}+\cdots,
\end{equation}
where $\cdots$ stands for the corrections all of which vanish in the $M\to 0$ limit.  The dilation around $M=0$ makes the behavior of $Q(M(1-\delta))$ stationary within a wide region at $\delta$ close to $1$ and, as $\delta\to 1$, $Q(M(1-\delta))$ converges to $\log\sqrt{2}$ at any finite $M$.   

Then, we like to show that the delta expansion on the large $M$ series of $Q$ recovers the flatness and emerged plateau indicates the correct value, $\log \sqrt{2}$.  Figure 4 shows the plot of 
\begin{equation}
\fl D[Q]=\log 2-\frac{1}{4}\Big(\log M-\sum_{k=1}^{K}\frac{1}{k}\Big)-\Big(
\begin{array}{c}
K+1 \\
1
\end{array}
\Big)\frac{1}{M}+\Big(
\begin{array}{c}
K+2 \\
2
\end{array}
\Big)\frac{2}{M^2}+\cdots\nonumber
\end{equation}
at 2nd, 5th, 8th, 11th, 14th and 17th orders.    At odd orders there exists only one extremum value and it may be regarded as the representative values on the plateau.  Then, as proposed in \cite{pms}, it is natural to take the extremum value as the approximation of $\log A=\log\sqrt{2}=0.346574\cdots$.   The results at $5$th, $11$th and $17$th orders are
\begin{equation}
\log A:\quad 0.3098, \quad 0.3397,\quad 0.3450
\end{equation}
and these values occur at $M=13.4558,\,15.7481,\,16.7$, respectively.  Our results above several orders are found to be in good agreement with the exact value.  The mass gap can be computed by using above approximants for $\log A$ by $m=(A/2)^{2/3}\lambda^{1/3}$, though we omit explicit results.

\begin{figure}[h]
\begin{center}
\includegraphics[scale=0.7]{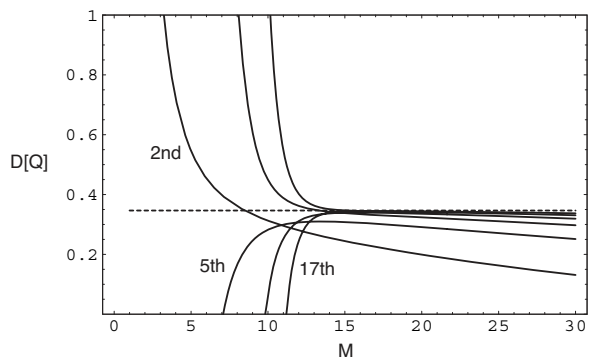}
\end{center}
\caption{Plots of $D[Q]=D[ \log \beta+3/4\log M]$ at $2$nd, $5$th, $8$th, $11$th, $14$th and $17$th orders.   The dotted line represents $\log \sqrt{2}=0.346574\cdots$.}
\end{figure} 

\subsection{Other truncations in square schemes}
In the previous subsection, a square truncation was investigated under the condition $N=K$.  There are other infinite choices of the truncation rule in the class of square schemes.  We here study the results of choices that 
\begin{equation}
N=K+L,
\end{equation}
where $L$ will be assigned some positive integer.  
We write here the result of evaluating the amplitude of scaling, $\log \sqrt{2}$, which is the leading term of $Q$ at small $M$.   
The results are summarized in Table 1.

\begin{table}
\caption{Approximation of the scaling amplitude $\log \sqrt{2}=0.346574\cdots$in various truncations in the square delta expansion.   The value of $M$ at which $Q$ reaches extremum is written in the parenthesis.}
\begin{indented}
\lineup
\item[]
\begin{tabular}{lllll}
\br
order $K$ & $N=K$ & $N=K+1$ & $N=K+2$ & $N=K+3$\\
\noalign{\smallskip}\hline\noalign{\smallskip}
5 & 0.30985 (13.456) & 0.31341 (15.305) &  0.31601 (17.152)&0.31799 (18.996)  \\
11 &0.33974 (15.748) & 0.34029 (16.783) & 0.34075 (17.816) &0.34113 (18.848) \\
17 &0.34501 (16.700) & 0.34513 (17.420) & 0.34523 (18.140) &0.34532 (18.859)\\
\br
\end{tabular}
\end{indented}
\vskip 0.2cm
\begin{indented}
\lineup
\item[]\begin{tabular}{lllll}
\br
order $K$ & $N=K+4$ & $N=K+5$  & $N=K+10$ & $N=K+20$ \\
\noalign{\smallskip}\hline\noalign{\smallskip}
5 & 0.31954 (20.839) & 0.32079 (22.680) & 0.32457 (31.874) & 0.32760 (50.239) \\
11 & 0.34145 (19.879) & 0.34173 (20.910)& 0.34267 (26.056) & 0.34356 (36.328) \\
17 & 0.34539 (19.578) & 0.34546 (20.296) & 0.34569 (23.883) & 0.34594 (31.041)\\
\br
\end{tabular}
\end{indented}
\vskip 0.2cm
\begin{indented}
\lineup
\item[]\begin{tabular}{lll}
\br
order $K$ & $N=K+30$ & $N=K+40$  \\
\noalign{\smallskip}\hline\noalign{\smallskip}
5 & 0.32891 (68.593) & 0.32963 (86.943)\\
11 & 0.34397 (46.587) &0.34421 (56.840) \\
17 & 0.34606 (38.189) & 0.34613 (45.331)\\
\br
\end{tabular}
\end{indented}
\end{table}

We find that as $L$ increases the approximant increases monotonically.  Further, approximation is improved as $L$ increases.   Note that the value of $M$ at which the function $D[Q]$ becomes extremum also grows with $N$.  Let the value of $M$ at which $D[Q]$ takes extremum value be $M^{*}$.  Then, at $K=17$, the ratio $M^{*}/N$ has the following values at $L=0,1,2,\cdots,5,10,20,30$ and $40$,
\begin{eqnarray}
& &0.982355, 0.967794, 0.954736, 0.942958, 0.932281,\nonumber\\
& &0.922556, 0.884561, 0.838967, 0.812543, 0.795287.
\end{eqnarray}
The ratio is of order $O(1)$ and decreases gradually.  Now, it is apparent that, in small $M$ region, the delta expanded $1/M$ series cannot offer us the scaling.  Rather, scaling can be observed at the region of $M\sim O(N)$.  Hence the small $M$ region can be neglected and the correlated limit (\ref{limit}) turns out to be a natural limit to consider.  This is the reason behind considering BT limit.

We emphasize that scaling reveals itself at $M\sim O(N)$ is a characteristic feature of square schemes.  In the conventional scheme \cite{yam,hhy}, the scaling is observed at $M\sim O(1)$.

\subsection{Results in Borel transform limit}
In this subsection, we study the scaling behavior of $\beta$ via $1/M$ expansion by the use of Borel transform.  

The Borel transform of $\log\beta$ reads at large $\hat M$ 
\begin{equation}
\fl B[\log\beta]=\log \frac{2}{\hat M}+\gamma_{E}-\frac{1}{\hat M}+\frac{2}{2!\hat M^2}+\cdots=\log\frac{2}{\hat M}+\gamma_{E}+\sum_{n=1}^{\infty}\frac{b'_{n}}{n! \hat M^{n}}.
\label{borel1}
\end{equation}
The behavior of (\ref{borel1}) is plotted in Figure 5 at orders $2$nd, $3$rd, $\cdots$ and $8$th.  It is obvious from (\ref{borel1}) that Borel transform improves the small $\hat M$ behavior:  Comparing (\ref{borel1}) with(\ref{30}), $B[\log \beta]$ has the inverse factorial in the coefficient and its convergence radius is larger.   In the model under consideration, $\log\beta$ converges for $0\le M^{-1}\le \frac{1}{4}$.  Hence $B[\log\beta]$ has infinite radius of convergence.  Thus the scaling behavior may be seen in $1/\hat M$ series.

\begin{figure}[h]
\begin{center}
\includegraphics[scale=0.7]{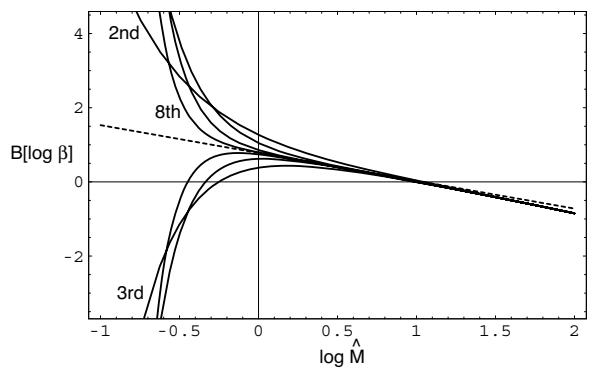}
\end{center}
\caption{Plots of $B[\log \beta]$ from 2nd to 8th orders.  The dotted line represents $\log[\sqrt{2}\hat M^{-3/4} /\Gamma(3/4+1)]$.}
\end{figure} 
The evaluation of the exponent $\alpha$ is easily done by considering the Borel transform of $\frac{\partial \log \beta}{\partial \log M}$.  We here omit the task and directly turn to the evaluation of the amplitude $A$ which is connected to the magnitude of mass gap.  We consider $Q=\log \beta+\frac{3}{4}\log M$ and its Borel transform.   Figure 6 shows the plots of $B[Q]=\log 2-\frac{1}{4}(\log\hat M-\gamma_{E})+\sum_{n=1}^{\infty}\frac{b'_{n}}{n!} \hat M^{-n}$ and $B[Q]|_{\hat M\to 0}=\log \sqrt{2}$.  

\begin{figure}[h]
\begin{center}
\includegraphics[scale=0.7]{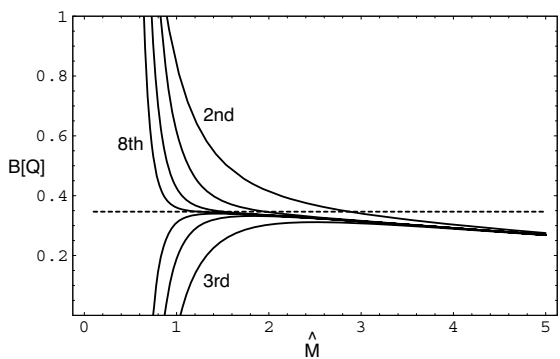}
\end{center}
\caption{Plots of $B[Q]$ from 2nd to 8th orders.  The dotted line represents $\log\sqrt{2}$.}
\end{figure} 

In terms of $\hat M$, the transformed series does not possess a plateau of wide region.  This is a natural result since values of $M$ of $O(N)$ correspond to values of $\hat M$ of $O(1)$.  In other words a wide plateau at $M\sim O(N)$ in square schemes are mapped to a region of $\hat M\sim O(1)$.  Since the stationarity is conserved by the mapping, however, we approximate $\log A$ by stationary value of $B[Q]$ .  The results at orders $5$th, $11$th and $17$th are as follows:
\begin{equation}
\log A:\quad 0.332143,\quad 0.345087,\quad 0.346381.
\end{equation}
These values are realized at $\hat M=1.834, 1.024, 0.712$, respectively.  It is now sure that the above sequence approaches to $\log\sqrt{2}=0.346574\cdots$, the exact value of the amplitude.  The corresponding mass gap is computed as $m/\lambda^{1/3}=0.7861,\,0.79291,\, 0.79359$ and in good agreement with the exact value, $m/\lambda^{1/3}=2^{-1/3}=0.793701\cdots$.   Comparison of the above approximants to those in square schemes at finite $L$ clarifies that BT limit produces best results.

\section{Application to other models}
Also in models we are going to discuss from now on, BT limit in a class of square schemes gives most accurate result than any of finite $N$.  The reason would be that the power like correction $M^{k}$ $(k>0)$ remains to $k-1$th order square scheme but, in BT limit, it vanishes at every order.  This means that the correction to the asymptotic scaling is smaller in BT limit.  Thus, we report the results only in BT limit.
\subsection{Anharmonic oscillator}
In this subsection we apply the delta expansion in the BT limit to the single component anharmonic oscillator.  At ${\cal N}=1$, the anharmonic oscillator cannot be solved and only numerical results are known to high accuracy.  It therefore serves us a good theoretical laboratory to examine our method.

In terms of the rescaled field $\varphi_{n}$ on a site $n$, the action $S$ reads
\begin{equation}
S=\sum_{n} V(\varphi_{n})-\beta\sum_{n} \varphi_{n+1}\varphi_{n},\quad V(\varphi)=\beta \varphi^2+\varphi^4.
\end{equation}
Even when the system is massless at the level of action, the fields $\varphi_{0}$ and $\varphi_{n}$ at large separation has the finite correlation length.  Then, Fourier transform of the two point function $\la\varphi_{0}\varphi_{n}\ra$ at large $n$ defines $M$ corresponding to (\ref{mass}).  By calculating $\la\varphi_{0}\varphi_{n}\ra$ by the use of hopping expansion \cite{hop}, one can obtain $M$ as a series in $\beta$.  Then, inverting $M^{-1}$ and $\beta$, we have large mass expansion of $\beta$.  In \cite{hhy}, $\beta$ is obtained up to $M^{-9}$.  Unfortunately, as in the large {\cal N} case, the Borel transform of $\beta(M)$ does not show clear sign of the continuum scaling in its effective region even at $9$th order.  However, taking the logarithm of $\beta$ improves the state of affairs.  We find
\begin{eqnarray}
\fl \log\beta=-\log(\rho M)+\Big(-3+\frac{1}{4\rho^2}\Big)\frac{1}{M}+\Big(6-\frac{5}{4\rho^2}+\frac{1}{12\rho^4}\Big)\frac{1}{M^2}\nonumber\\
 +\Big(-\frac{33}{2}+\frac{47}{8\rho^2}-\frac{27}{32\rho^4}+\frac{5}{128\rho^6}\Big)\frac{1}{M^3}+\Big(\frac{411}{8}-\frac{431}{16\rho^2}+\frac{983}{160\rho^4}-\frac{457}{768\rho^6}\nonumber\\
 +\frac{385}{18432\rho^8}\Big)\frac{1}{M^4}
 +\Big(-\frac{3417}{20}+\frac{973}{8\rho^2}-\frac{1553}{40\rho^4}+\frac{703}{120\rho^6}-\frac{1303}{3072\rho^8}+\frac{61}{5120\rho^{10}}\Big)\nonumber\\ \times \frac{1}{M^5}
 +\Big(\frac{2367}{4}-\frac{2173}{4\rho^2}+\frac{72411}{320\rho^4}-\frac{45199}{960\rho^6}+\frac{47081}{8960\rho^8}-\frac{1391}{4608\rho^{10}}\nonumber\\
 +\frac{583}{82944\rho^{12}}\Big)\frac{1}{M^6}
 +\Big(-\frac{117981}{56}+\frac{38511}{16\rho^2}-\frac{800529}{640\rho^4}+\frac{85869}{256\rho^6}-\frac{3667987}{71680\rho^8}\nonumber\\
  -\frac{387119}{86016\rho^{10}}-\frac{62843}{294912\rho^{12}}
 +\frac{52195}{12386304\rho^{14}}\Big)\frac{1}{M^7}
 +\Big(\frac{489681}{64}-\frac{678201}{64\rho^2}\nonumber\\
 +\frac{3409315}{512\rho^4}-\frac{11285453}{5120\rho^6}+\frac{1387540351}{3225600\rho^8}-\frac{7581293}{147456\rho^{10}}+\frac{10207705}{2752512\rho^{12}}\nonumber\\
 -\frac{350285}{2359296\rho^{14}}+\frac{861575}{339738624\rho^{16}}\Big)\frac{1}{M^8}+O(M^{-9})\nonumber\\
=-\log(\rho M)+\sum_{k=1}^{\infty}\frac{b'_{k}}{M^k},
\end{eqnarray}
where $\rho=\Gamma(3/4)/\Gamma(1/4)=0.337989\cdots$.  Then at large $M$, Borel transform of $\log\beta$ is given by
\begin{equation}
B[\log\beta]=-\log\rho-\log \hat M+\gamma_{E}+\sum_{k=1}^{\infty}\frac{b'_{k}}{k!} \hat M^{-k}.
\end{equation}
Here, truncation of $B[\log\beta]$ up to $\hat M^{-K}$ will be called $K$th order approximant.  
Now, the scaling behavior of $\log\beta$ is given by
\begin{equation}
\log\beta\sim \log A-\frac{3}{4}\log M,
\end{equation}
and its Borel transform reads
\begin{equation}
B[\log\beta]\sim \log A-\frac{3}{4}(\log \hat M-\gamma_{E}),
\end{equation}
where $\log A$ is known numerically as $\log A=0.81841\cdots$ {\footnote {The value of $\log A$ is calculated from the value of the mass gap referred in \cite{bel} }}.

Figure 7 shows the plots of $B[\log \beta]$ at $1$st and $8$th orders.  At $8$th order, we find that the scaling behavior is observed around $\log \hat M\sim 0$ in the Borel transformed $1/M$ series.
\begin{figure}[h]
\begin{center}
\includegraphics[scale=0.7]{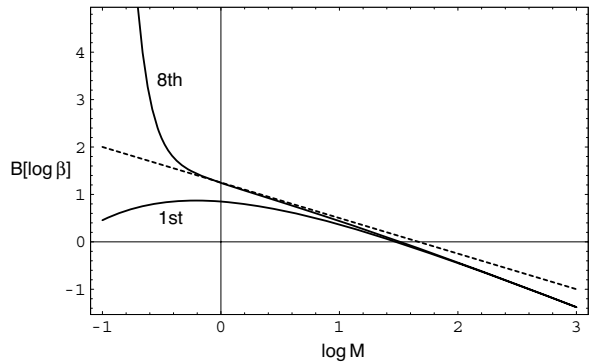}
\end{center}
\caption{$B[\log\beta]$ at $1$st and $8$th orders.  The dotted line represents $\log A-\frac{3}{4}(\log \hat M-\gamma_{E})$ where $\log A=0.81841\cdots$.}
\end{figure} 
Having captured the scaling behavior, we can evaluate the constant $A$ which leads to the mass gap in the continuum limit by $m=(A/2)^{2/3}\lambda^{1/3}$.  The evaluation step goes as in the previous section.  We deal with $Q=\log\beta+\frac{3}{4}\log M$ and consider its Borel transform.  $B[Q]$ is plotted in Figure 8.   We find that the effective region grows to the smaller $\hat M$ region as the order increases.  The value of $\log A$ is indicated by the stationary value as in the case of large ${\cal N}$ limit.  At $K=1, 3, 5, 7$, we have following approximants of $\log A$:
\begin{equation}
\log A:\quad 0.68467,\quad 0.75734,\quad 0.78197,\quad 0.79358.
\end{equation}
Since the exact value of $\log A$ is $0.81841\cdots$ we can say that delta expansion in BT limit is successfully working.  The mass gap $m$ is then computed at respective orders as follows:
\begin{equation}
\frac{m}{\lambda^{1/3}}:\quad 0.9943,\quad 1.0437,\quad 1.0610,\quad 1.0692.
\end{equation}
The exact value of $m$ is known to be $m=1.087096 \cdots\times \lambda^{1/3}$ \cite{bel} and the results at several  orders give good approximation.

\begin{figure}[h]
\begin{center}
\includegraphics[scale=0.7]{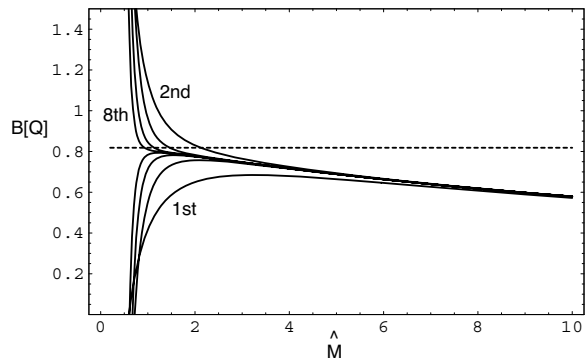}
\end{center}
\caption{Plots of $B[Q]=B[\log\beta+\frac{3}{4}\log M]$ at $K=1, \cdots, 8$.  The dotted line represents the value, $\log A=0.81841\cdots$.}
\end{figure}

\subsection{2d non-linear $\sigma$ model}
Up to now, we considered quantum mechanical cases where the models are defined on one-dimensional lattice.  In this subsection, we discuss a model field theory, the non-linear $\sigma$ model in the large ${\cal N}$ limit at two dimension.   The 2d non-linear $\sigma$ model can be solved out in the large ${\cal N}$ limit and enjoys interesting properties of asymptotic freedom and dynamical mass generation.  We study the continuum scaling behavior via the large mass expansion under BT limit.

The 2d non-linear $\s$ model on continuum Euclidean space is defined by the action, $
{\cal L}={1 \over 2f}\sum_{\mu}\big(\partial_{\mu} \vec\s\big)^2$ 
where $f$ denotes the bare coupling constant and the fields $\s^{A}(x)\,\,(A=1,2,\cdots,{\cal N})$ obey the constraint, $\vec{\s}^2(x)=\sum_{A=1}^{\cal N}\s^{A}(x)\s^{A}(x)={\cal N}$.  
The discretized space we work with is the periodic square lattice with the lattice spacing
$a$ where a site is labeled by two integers, $(n_{1},n_{2})={\bf n}$.  A simple version of the lattice action is given by
\begin{equation}
S=2\beta \sum_{\bf n}\vec{\s}_{\bf n}^2-\beta\sum_{\bf n}\sum_{\mu=1,2}\vec{\s}_{\bf n}\cdot\vec{\s}_{\bf n+\bf e_{\mu}},
\label{sigmaaction}
\end{equation}
where
\begin{equation}
\beta:=\frac{1}{f}
\end{equation}
and we call $\beta$ as the hopping parameter.  In (\ref{sigmaaction}), 
$\vec{\s}_{\bf n+\bf e_{\mu}}$ stands for the nearest neighbour spin of $\vec{\s}_{\bf n}$ with ${\bf e}_{1}=(1,0)$ and ${\bf e}_{2}=(0,1)$.  The constraint is written as 
$\vec{\s}_{\bf n}^2={\cal N}$ and the first term in (\ref{sigmaaction}) is actually a constant that can be omitted. 

In the large ${\cal N}$ limit, the model can be solved out and $\beta$ is given from the constraint as
\begin{equation}
\beta=\int^{\pi}_{-\pi}\frac{d^{2}{\bf p}}{(2\pi)^2}\frac{1}{M+2\sum_{\mu=1,2}(1-\cos p_{\mu})},
\label{closedbeta}
\end{equation} 
where
\begin{equation}
M=(ma)^2.
\end{equation}
We start our analysis with $1/M$ expansion of $\beta$, which is equally available in other lattice models.  
 It is easy from (\ref{closedbeta}) to obtain
\begin{equation}
\beta={1 \over M}-{4 \over M^2}+{20 \over
M^3}-{112
\over M^4}+{676 \over M^5}-{4304
\over M^6}+O(M^{-7}).
\label{betastrong}
\end{equation}

Previous two models have a common feature that the logarithm of $\beta$ is more convenient to address the continuum scaling than $\beta$ itself.  It does not apply to the present model since $\beta$ itself behaves logarithmically in the scaling region, $\beta\sim -\frac{1}{4\pi}\log(M/32)$.  $\log\beta$ behaves like $\sim \log(-\frac{1}{4\pi}\log[M/32])$ at small enough $M$, and to capture such a behavior is more difficult.   Hence, we consider  Borel transform of $\beta$ which is easily obtained as 
$
B[\beta]=\sum_{k=1}^{\infty}\frac{b_{k}}{n! \hat M^k}$ at $\hat M\gg 1$ and at the scaling region, 
$B[\beta]\sim -\frac{1}{4\pi}(\log\hat M-\gamma_{E}-\log 32)\, (\hat M\ll 1)$.   Unfortunately, the scaling behavior of $B[\beta]$ is seen at $\hat M\gg 1$ only roughly, though remarkable improvement is found as shown in Figure 9(2).   To improve the status, we employ Symanzik improvement of the lattice action \cite{sym}.   The reason why Symanzik's modification of the lattice action helps us to capture the asymptotic scaling is described in detail in \cite{yam}.  To say it briefly, the leading correction to the asymptotic scaling is $M\log M$ and this affects the small $\hat M$ behavior even at small $\hat M$.  Then, the term can be subtracted by introducing the next-to-the-nearest neighbour coupling term $\sum_{\mu}\vec{\s}_{\bf n}\cdot\vec{\s}_{{\bf n}+2{\bf e}_\mu}$ into the action.  At the first order of Symanzik imporovement program, the action becomes
\begin{figure}[h]
\begin{center}
\includegraphics[scale=0.8]{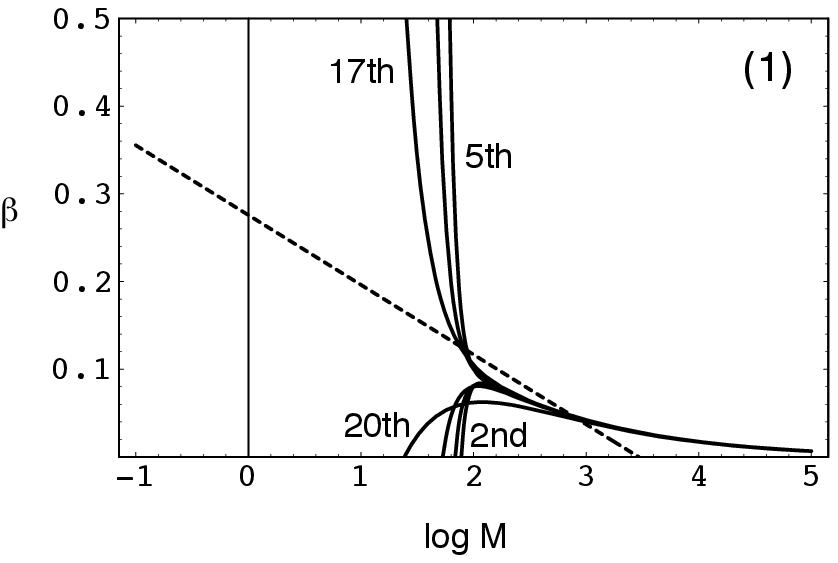}
\includegraphics[scale=0.8]{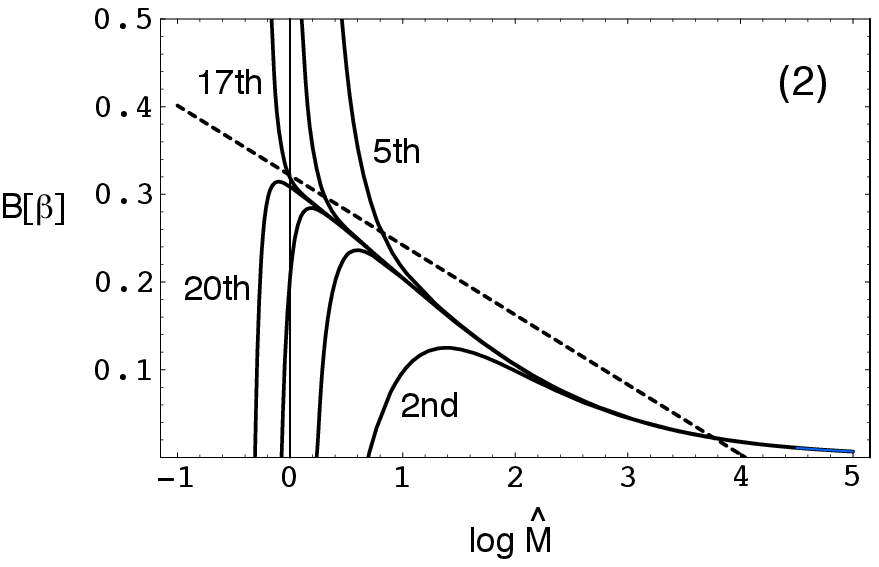}
\includegraphics[scale=0.8]{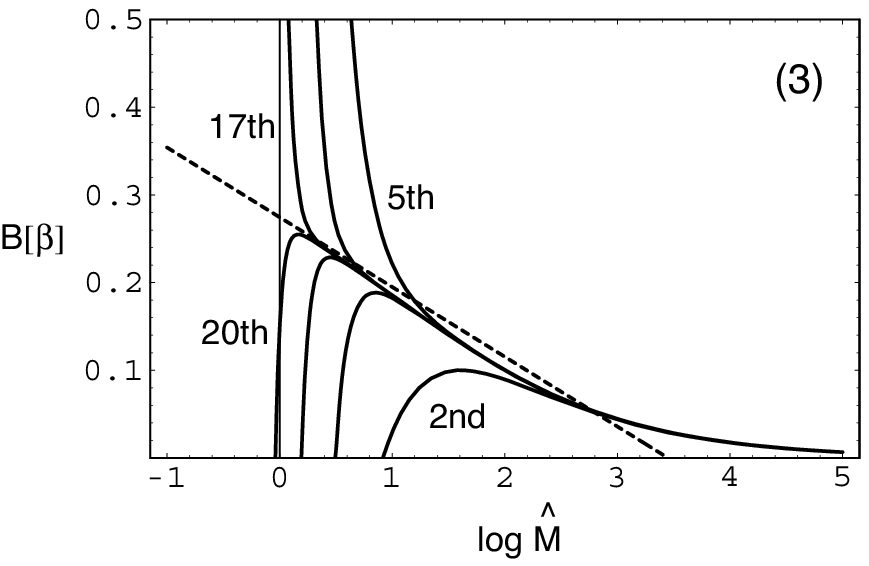}
\end{center}
\caption{(1) Plots of $\beta$ in the large $M$ expansion at $2$nd, $5$th, $8$th, $11$th, $14$th, $17$th and $20$th orders.  (2) Plots of $B[\beta]$ in the large $M$ expansion at $2$nd, $5$th, $8$th, $11$th, $14$th, $17$th and $20$th orders.  (3) Plots of $B[\beta]$ in the large $M$ expansion at $2$nd, $5$th, $8$th, $11$th, $14$th, $17$th and $20$th orders at the first order Symanzik program. }
\end{figure}    

\begin{equation}
S=\beta\sum_{\bf n}
\bigg[\frac{5}{2}\vec{\s}_{\bf n}^2-\frac{4}{3}\sum_{\mu}\vec{\s}_{\bf n}\cdot\vec{\s}_{{\bf n}+{\bf e}_\mu}+\frac{1}{12}
\sum_{\mu}\vec{\s}_{\bf n}\cdot\vec{\s}_{{\bf n}+2{\bf e}_\mu}\bigg].
\end{equation}
The constraint relation at large ${\cal N}$ now reads
\begin{equation}
\beta=\int_{-\pi}^{\pi}\frac{d^2 {\bf p}}{(2\pi)^2} \frac{1}{
M+5-\frac{8}{3}\sum_{\mu}\cos p_{\mu}+\frac{1}{6}\sum_{\mu}\cos 2p_{\mu} }.
\label{sym}
\end{equation}
In the improved action, the large $M$ expansion and the value of the constant part in the scaling behavior becomes as follows:  Large $M$ expansion reads \cite{yam},
\begin{equation}
\beta=\frac{1}{M}-\frac{5}{M^2}+\frac{1157}{36M^3}-\frac{8419}{36M^4}+\cdots=\sum_{k=1}^{\infty}\frac{b_{k}}{M^k}
\end{equation}
and the scaling behavior,
\begin{equation}
\beta \sim -\frac{1}{4\pi}\log M+B',
\end{equation}
where $B'=B-0.0471699\cdots=0.2286245\cdots$.  By plotting $B[\beta]$ in $1/\hat M$ expansion, we find clear logarithmic scaling in accordance with the asymptotic freedom (see Figure 9(3)).

Having observed the scaling, we turn to the evaluation of the constant $B'$ which is directly connected to the mass gap by $m=\exp(2\pi B') \Lambda$ ($\Lambda$ stands for the scale parameter in the first order Symanzik model).   For the purpose it is convenient to deal with $Q=\beta+\frac{1}{4\pi}\log M=\frac{1}{4\pi}\log M+\sum_{k=1}^{\infty}\frac{b_{k}}{M^k}$ which tends to $B'$ in the $M\to 0$ limit.  By Borel transforming $Q$ we have plotted the resulting functions at $2$nd, $5$th, $8$th, $11$th, $14$th and $17$th orders (see Figure 10).
\begin{figure}[h]
\begin{center}
\includegraphics[scale=0.8]{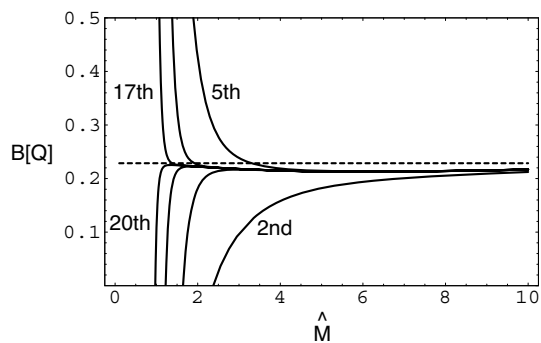}
\end{center}
\caption{Plots of $B[Q]=B[\beta+\frac{1}{4\pi}\log M]$ at $K=2,5,8,11,14,17.20$.  The dotted line represents the value of $B'=0.2286245\cdots$.}
\end{figure} 
Approximation of $B'$ by extremum values are shown in Table 2.   In BT limit, there appear two extrema at 6th order.  This is a signal of appearance of new family of extrema at higher orders.  First family of extrema stays at $\hat M\sim 6$ to all higher orders.  The second family appears from 6th order and the location moves to smaller $\hat M$ as the order increases.  It is obvious that we should trace the second family, since it signals $\lim_{\hat M\to 0}B[Q]$.

To summarize, the sequence of extrema at smallest $\hat M$ suggests strongly that it converges to the exact value of $B'$ and the dynamical mass can be calculated from the large $M$ expansion of $\beta$.

\begin{table}
\caption{Approximants of $B'$ in Borel transform (BT) limit.  Exact value of $B'$ is $B'=0.2286245\cdots$.  The value of $\hat M$ at which $B[Q]$ becomes extremum is shown in the parenthesis.}
\vskip 5pt
\label{tab:2}       
\begin{indented}
\lineup
\item[]\begin{tabular}{lll}
\br
order &  $B'$ approximants  &  $B'$ approximants\\
\noalign{\smallskip}\hline\noalign{\smallskip}
2  &  &\\
3  & 0.215929 (7.836) & \\
4  &   & \\
5 & 0.213059 (6.372) & \\
6 &  0.212669 (5.915) &  0.213555 (3.961) \\
7 &   0.212765 (6.066) & \\
8 &   0.212748 (6.038)  &  0.216666 (2.966) \\
9 & 0.212751 (6.043) & \\
10 & 0.212751 (6.042) & 0.219193 (2.436)\\
20 &  0.212751 (6.043)  & 0.225305 (1.340) \\
\br
\end{tabular}
\end{indented}
\end{table}

\subsection{2d Gross-Neveu model with Wilson fermion}
The models considered up to now allow both summation schemes, conventional and square ones to us for simulating continuum limit.  It would be nice if the truncation presented in this work has an essential importance for the purpose.

Here we show a theoretical example, $2d$ Gross-Neveu model in the large ${\cal N}$ limit \cite{gn} with Wilson fermion \cite{wil}, where conventional scheme fails but square and BT schemes succeed to capture the continuum scaling.

Gross-Neveu model describes quartic interactions between massless ${\cal N}$ component fermion fields $\psi_{\alpha}^{A}\,\,(A=1,2,\cdots,{\cal N};\,\alpha=1,2)$.  Introducing the auxiliary fields $\sigma$ on each site, the action on the lattice reads \cite{aok}
\begin{eqnarray}
S&=&-\frac{a}{2}\sum_{{\bf n},\mu}[\bar\psi({\bf n})(r-\gamma_{\mu})\psi({\bf n+\bf e_{\mu}})+\bar\psi({\bf n+\bf e_{\mu}})(r+\gamma_{\mu})\psi({\bf n})]\nonumber\\& &+\sum_{{\bf n}} (2ar+a^2\sigma({\bf n}))\bar\psi({\bf n})\psi({\bf n})+\frac{{\cal N}a^2}{2g^2}\sum_{{\bf n}}(\sigma({\bf n})-\delta m)^2,
\end{eqnarray}
where $\gamma$ matrices are given by Pauli matrices,
\begin{equation}
\gamma_{1}=\sigma_{2},\quad \gamma_{2}=\sigma_{1},\quad \gamma_{5}=\sigma_{3}=i\gamma_{1}\gamma_{2}.
\end{equation}
The parameter $r$ is called Wilson parameter and as long as $r\neq 0$, the $\gamma_{5}$ symmetry related to the transformation $\psi\to \gamma_{5}\psi$ is explicitly broken.  Thus the mass protection mechanism is absent and there exists mass divergence even when the current mass is zero.  The parameter $\delta m$ represents the contribution of the mass counter term to cancel out the divergence.  Now, to fix the mass counter term, we use the perturbative analysis and compute the $\sigma$ tad pole in the large ${\cal N}$ limit.  We employ the renormalization condition that the tad pole exactly vanishes.  Then we find
\begin{eqnarray}
\delta m&=&-\frac{2g^2}{a} C_{1}(r)\label{kurikomi}\\
C_{1}(r)&=&\int _{-\pi}^{\pi}\frac{d^2 p}{(2\pi)^2}\frac{r\sum_{\mu}(1-\cos p_{\mu})}{\{r\sum_{\mu}(1-\cos p_{\mu})\}^2+\sum_{\mu}\sin^2 p_{\mu}}.
\end{eqnarray}
Under this prescription, the mass cannot be created to all perturbative orders.  Hence, the dynamical mass can be generated only non-perturbatively.

One can discuss the recovery of the broken $\gamma_{5}$ symmetry in the continuum limit by studying the effective potential of $\sigma$.  In the large ${\cal N}$ limit, it is exactly calculated to give \cite{aok}
\begin{equation}
V_{L}=\frac{1}{2g^2}(\sigma_{L}-\delta m_{L})^2-\int^{\pi}_{-\pi}\frac{d^2p}{(2\pi)^2}\log\Big[\sum \sin^2p_{\mu}+(\sigma_{L}+r\sum(1-\cos p_{\mu}))^2\Big],
\label{epot}
\end{equation}
where $\sigma_{L}=\sigma a$, $\delta m_{L}=\delta m\cdot a$ and $V_{L}=V a^2$.  The restoration of $\gamma_{5}$ symmetry in $a\to 0$ limit is confirmed by expanding $V_{L}$ in  $\sigma_{L}$.  The result is
\begin{equation}
V_{L}=-\Big(\frac{\delta m_{L}}{g^2}+2C_{1}\Big)\sigma_{L}+\Big(\frac{1}{2g^2}-C_{0}'+2C_{2}\Big)\sigma^2_{L}+\frac{\sigma^2_{L}}{4\pi}\log\frac{\sigma^2_{L}}{e}+O(a^3),
\label{cont}
\end{equation}
where
$$
C_{0}'=\lim_{\sigma_{L}\to 0}\bigg\{\int_{-\pi}^{\pi}\frac{d^2 p}{(2\pi)^2}\frac{1}{\sum_{\mu}\sin ^2 p_{\mu}+\sigma_{L}^2+(r\sum_{\mu}(1-\cos p_{\mu}))^2}+\frac{\log \sigma_{L}^2}{4\pi}\bigg\}
$$
and
$$
C_{2}=\int_{-\pi}^{\pi}\frac{d^2 p}{(2\pi)^2} \bigg[\frac{2r\sum (1-\cos p_{\mu})}{\sum \sin^2 p_{\mu}+(r\sum (1-\cos p_{\mu}))^2}\bigg]^2.
$$

In (\ref{cont}), we discarded the leading constant that remains in the $\sigma_{L}\to 0$ limit, since it simply represents the uniform shift of vacuum energy.  
Note that the coefficient of $\sigma_{L}^2\log\sigma_{L}^2$ agees with the correct one and the doubling phenomenon is not occured.  
Under (\ref{kurikomi}), the linear term vanishes and thus the form of the effective potential agrees with that of the continuum limit,
\begin{equation}
V_{L}=\Big(\frac{1}{2g^2}-C_{0}'+2C_{2}\Big)\sigma^2_{L}+\frac{\sigma^2_{L}}{4\pi}\log\frac{\sigma^2_{L}}{e}+O(a^3).
\label{potlim}
\end{equation}
The dynamical mass of elementary fermion, $m_{D}$, is given by the solution of $\frac{\partial V_{L}}{\partial \sigma_{L}}=0$.  From (\ref{potlim}), we have the solution $
\sigma_{L}^2= \exp\big[4\pi(-\frac{1}{2g^2}+C_{0}'-2C_{2})\big]=(a m_{D})^2$.

Now, we turn to the issue that we have discussed through out this work.  We try to capture the continuum scaling in expansions effective at large lattice spacings.  In the case of the present model, we ask whether the large $\sigma_{L}(=\sigma a)$ expansion of the effective potential can be improved by the use of delta expansion or Borel transform.   From (\ref{epot}), $V_{L}$ at large $\sigma_{L}$ reads{\footnote {Though we have obtained $1/\sigma_{L}$ expansion from (\ref{epot}) which holds only in the large ${\cal N}$ limit, we stress that the expansion is generally available through the hopping expansion.  Here, we have used (\ref{epot}) only for the sake of efficiency.}}
\begin{eqnarray*}
V_{L}&=&\frac{1}{2g^2}(\sigma_{L}-\delta m_{L})^2-\bigg\{\log \sigma_{L}^2+\frac{4r}{\sigma_{L}}+\frac{1-5r^2}{\sigma_{L}^2}+\Big(-4r+\frac{28r^3}{3}\Big)\frac{1}{\sigma_{L}^3}\\
& &+\Big(-\frac{5}{8}+\frac{57r^2}{4}-\frac{169r^4}{8}\Big)\frac{1}{\sigma_{L}^4}+\Big(5r-50r^3+\frac{269r^5}{5}\Big)\frac{1}{\sigma_{L}^5}\\
& &+\Big(\frac{7}{12}-\frac{115r^2}{4}+\frac{705r^4}{4}-\frac{1781r^6}{12}\Big)\frac{1}{\sigma_{L}^6}\\
& &+\Big(-7r+145r^3-627r^5+\frac{3035r^7}{7}\Big)\frac{1}{\sigma_{L}^7}\\
& &+\Big(-\frac{169}{256}+\frac{3521r^2}{64}-\frac{87395r^4}{128}+\frac{144137r^6}{64}-\frac{338377r^8}{256}\Big)\frac{1}{\sigma_{L}^8}\\
& &+\Big(\frac{169r}{16}-\frac{4291r^3}{12}+\frac{24675r^5}{8}-\frac{32649r^7}{4}+\frac{599569r^9}{144}\Big)\frac{1}{\sigma_{L}^9}+\cdots\bigg\}.
\end{eqnarray*}

Fig. 11 shows the plot of $V_{L}\sigma_{L}^{-2}$ at $r=1/4,1/2,1$ to the several orders in $\sigma_{L}^{-1}$.
\begin{figure}[htb]
\begin{center}
\includegraphics[scale=0.7]{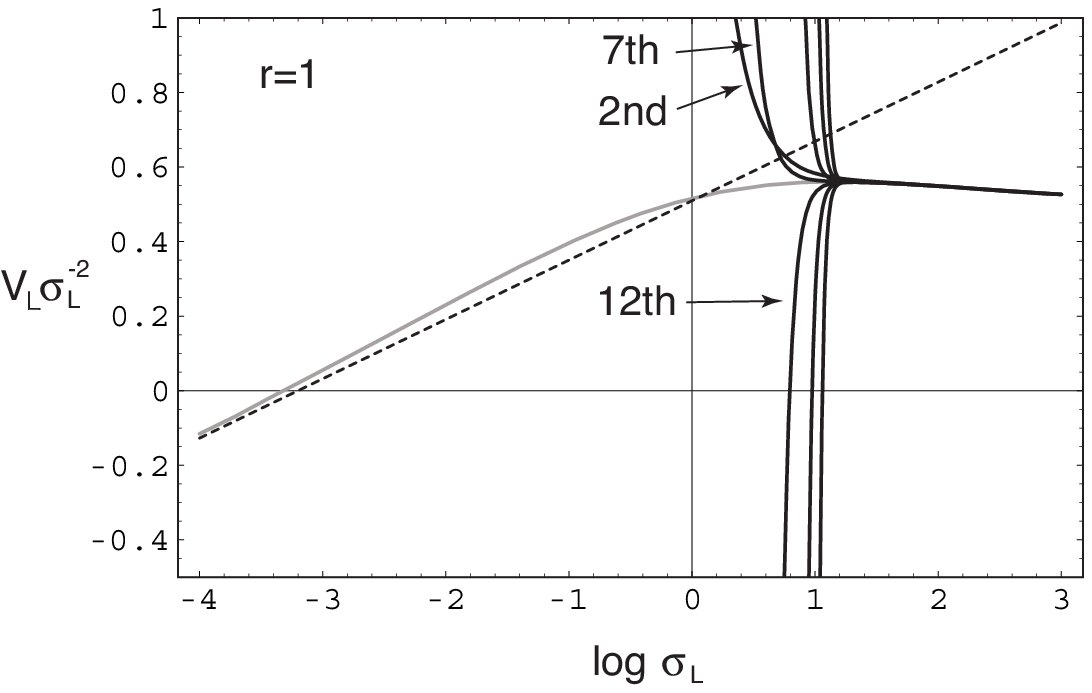}
\includegraphics[scale=0.7]{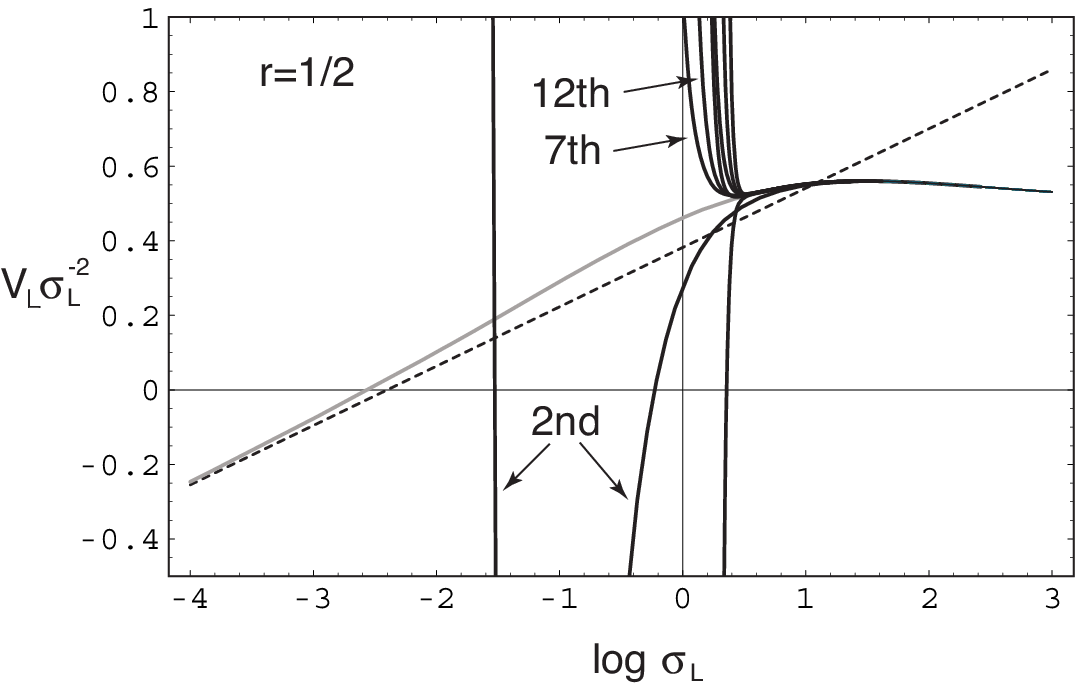}
\includegraphics[scale=0.7]{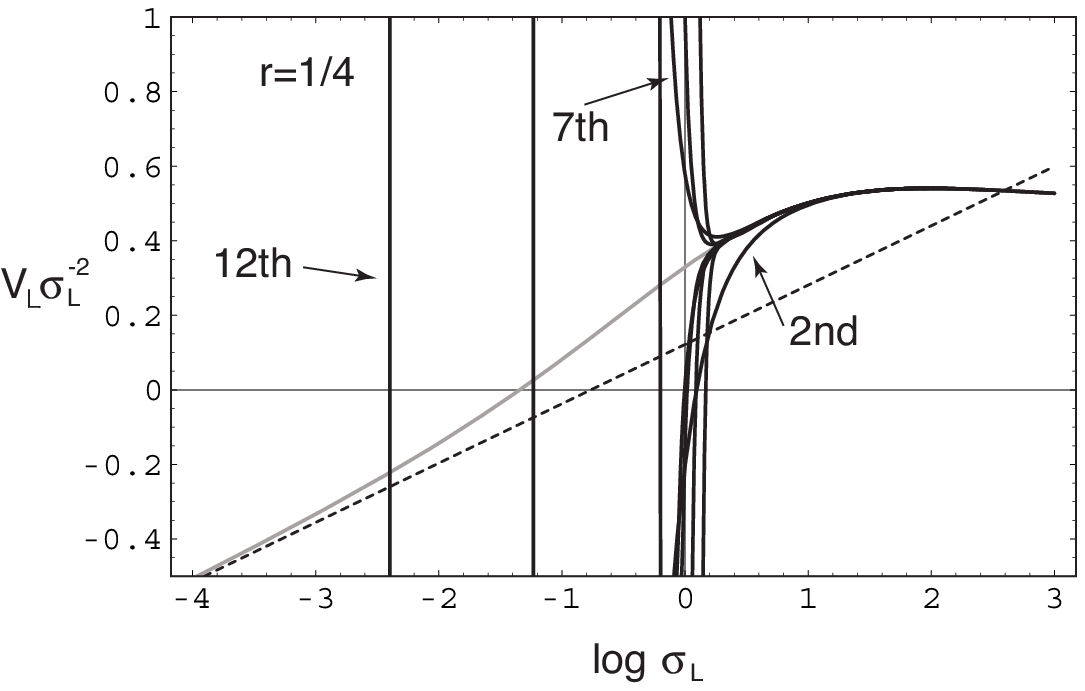}
\caption{Plots of the effective potential $V_{L}/\sigma_{L}^{-2}$ with Wilson fermion at $g=1$ and  $r=1/4,1/2,1$.  The solid black lines represent the potential expanded in $1/\sigma_{L}$ at 2nd, 7th,12th, $\cdots$ and 37th orders.  The solid gray line represents the exact potential and the dotted line the asymptotic behavior (\ref{potlim}) in the continuum limit.  At $r=1/4$, the hopping series is effective to $\log \sigma_{L}\sim 0$, while at $r=1$, effective to $\log\sigma_{L}\sim 1$.}
\end{center}
\end{figure}
The dotted lines represent the asymptotic scaling of $V_{L}\sigma_{L}^{-2}$ (see (\ref{potlim}))  in the continuum limit.  At $r=1$, which is almost the standard choice in the literatures, the deviation of the exact function to the asymptotic behavior is small.  For every sampled values of $r$, the asymptotic scaling begins around $\log \sigma_{L}\sim -3$.

The delta expansion technique in the conventional scheme is not adequate in the present model.  The difficulty originates from the existence of the linear divergence.  The one-loop counter term cancels out the linear divergence involved in the integral, but the cancelation becomes perfect only when the integral is expanded in $\sigma_{L}$.  When the integral is expanded in $1/\sigma_{L}$, we cannot isolate the divergent piece and the conventional delta expansion makes the cancellation incomplete.  The square type, on the otherhand, is useful to cope with the cancellation because all contributions are expanded to the same order of $\delta$ in the scheme.  Then, it is found that BT limit produces best simulation as in the previous models.  Hence in the following, we report the result of applying BT limit to obtain the continuum limit of the effective potential.

Now consider the Borel transform of $V_{L}\sigma_{L}^{-2}$.  Using $B[\sigma_{L}^{-k}]=\hat \sigma_{L}^{-k}/k!$ and
\begin{equation}
B[\frac{1}{\sigma_{L}^2}\log \sigma_{L}^2]=\frac{1}{\hat \sigma_{L}^2}\Big(\frac{1}{2}\log\hat\sigma_{L}^2-\gamma_{E}+\frac{3}{2}\Big),
\end{equation}
we obtain
\begin{eqnarray}
B[V_{L}\sigma_{L}^{-2}]&=&\frac{1}{2g^2}-\frac{\delta m_{L}}{g^2}\frac{1}{\hat\sigma_{L}}-\bigg\{\frac{1}{\hat \sigma_{L}^2}\Big(\frac{1}{2}\log\hat\sigma_{L}^2-\gamma_{E}+\frac{3}{2}\Big)\nonumber\\
& &+\frac{4r}{3!\hat\sigma_{L}^3}+\frac{1-5r^2}{4!\hat\sigma_{L}^4}+\Big(-4r+\frac{28r^3}{3}\Big)\frac{1}{5!\hat\sigma_{L}^5}+\cdots\bigg\}.
\end{eqnarray}
Fig. 12 shows the plot of $B[V_{L}\sigma_{L}^{-2}]$ for various values of $r$ at orders $2,7,12,\cdots,37$.  
\begin{figure}[htb]
\begin{center}
\includegraphics[scale=0.65]{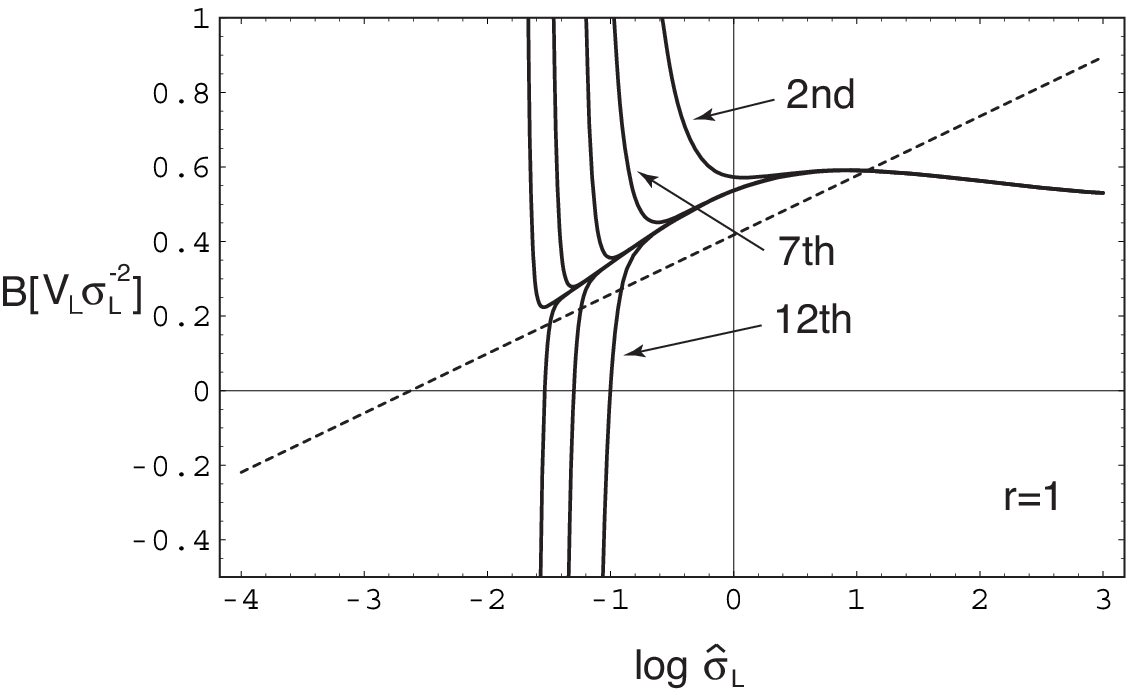}
\includegraphics[scale=0.65]{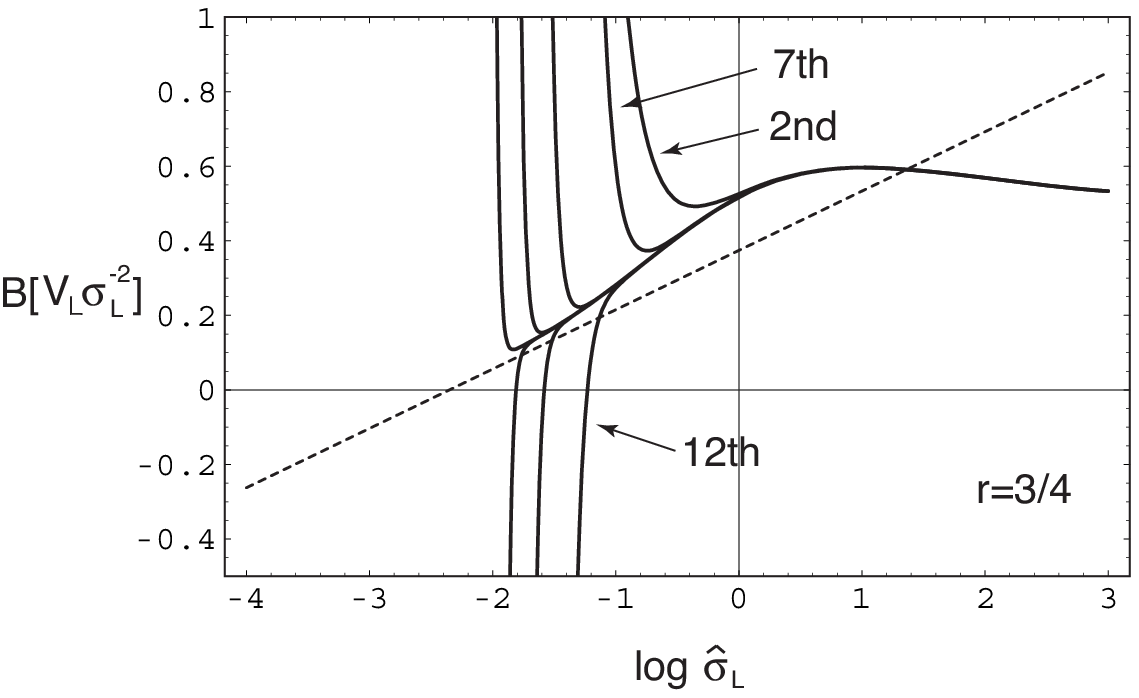}
\includegraphics[scale=0.65]{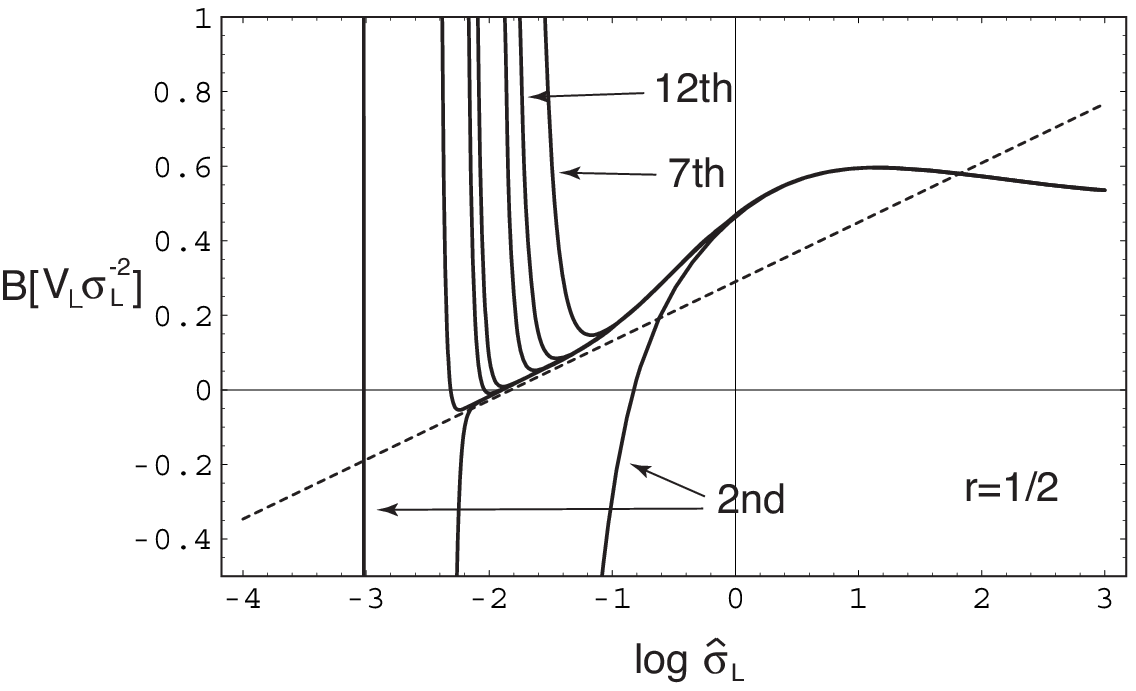}
\includegraphics[scale=0.65]{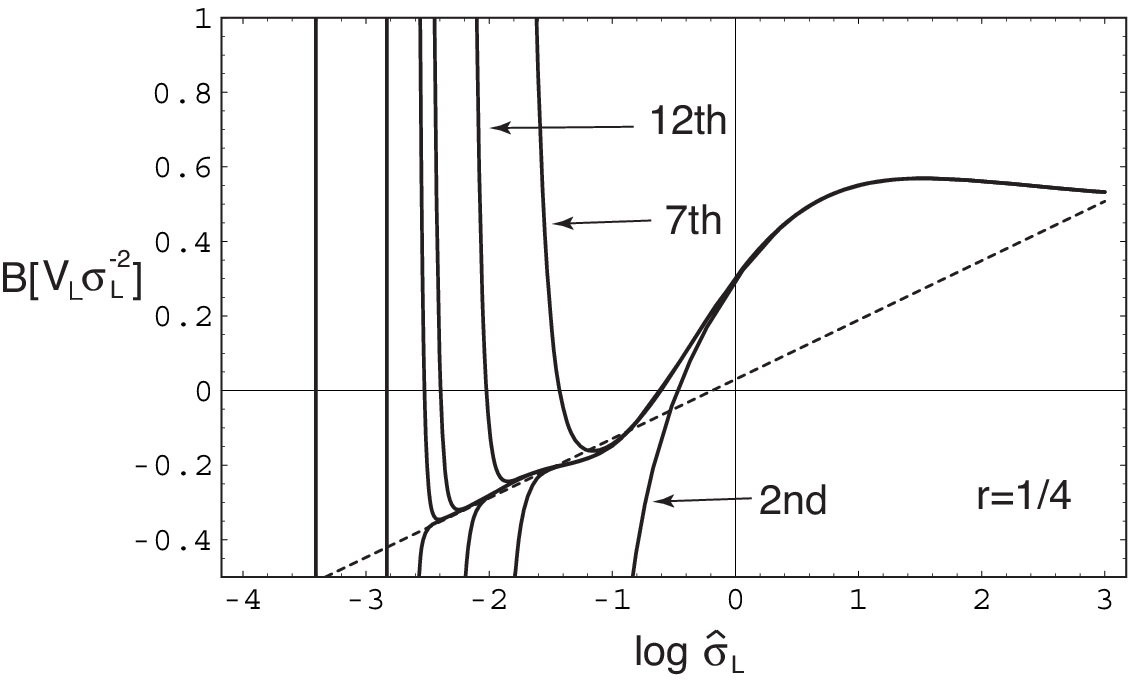}
\includegraphics[scale=0.65]{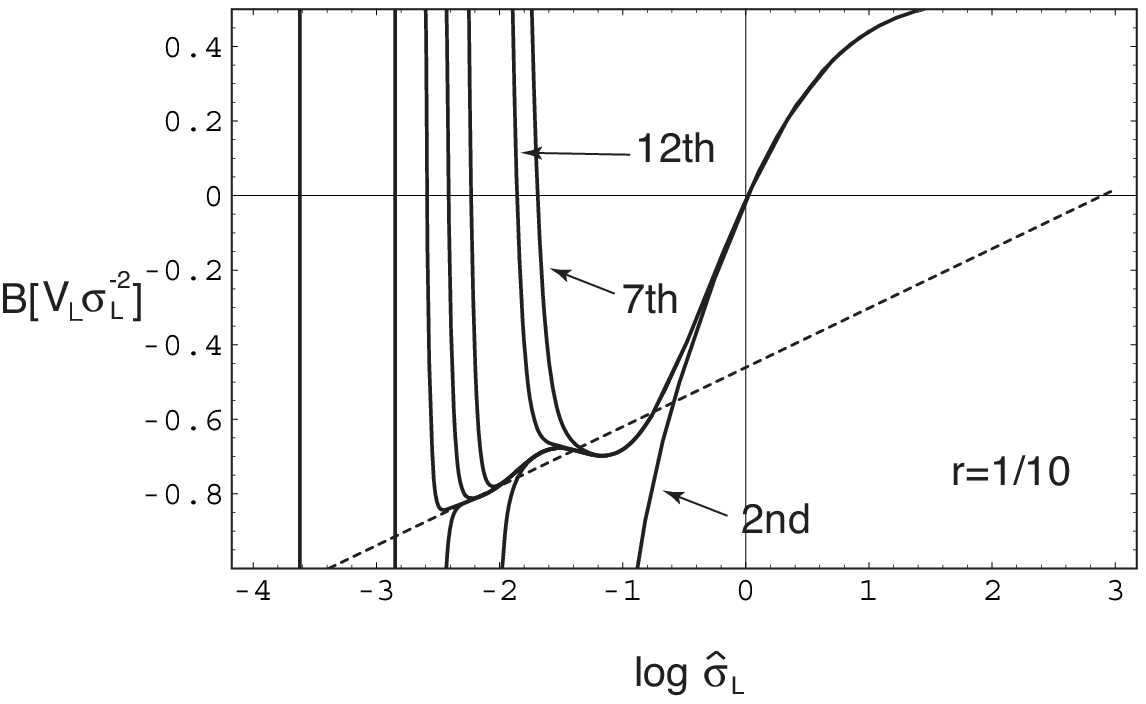}
\caption{Plots of Borel transformed large $\sigma_{L}$ expansion of $V_{L}\sigma_{L}^{-2}$ and its asymptotic scaling at $g=1$, $B[V_{L}\sigma_{L}^{-2}]\sim \frac{1}{2g^2}-C_{0}'+2C_{2}+\frac{1}{4\pi}(\log\frac{\sigma^2_{L}}{e}-2\gamma_{E})$ ($\gamma_{E}$: Euler constant).   From the first to the last, $r=1, 3/4,1/2,1/4,1/10$.  Dotted line represents the asymptotic scaling and the solid lines $1/\hat \sigma_{L}$ series at 2nd, 7th, 12th, $\cdots$ and 37th orders.}
\end{center}
\end{figure}
When $r=1/10$, the function oscillates at small $\hat\sigma_{L}$.  This property makes the quantitative analysis, for example the evaluation of $m_{D}$, difficult.  When $r=1/4$, the approach to the asymptotic scaling $B[V_{L}\sigma_{L}^{-2}]\sim \frac{1}{2g^2}-C_{0}'+2C_{2}+\frac{1}{4\pi}(\log\frac{\sigma^2_{L}}{e}-2\gamma_{E})$ is seen at 12th order and the oscillation is weak.  When $r=1/2$, asymptotic scaling is captured around $\log\hat \sigma_{L}\sim -1.5$ at 12th order and no oscillation is observed there.  In general, the $1/\hat \sigma_{L}$ series becomes effective to smaller $\hat \sigma_{L}$ for smaller $r$ but the expense is oscillatory behavior.  As found from (\ref{potlim}) the leading logarithmic correction is independent of $r$ but one cannot set $r=0$ from the outset.  In other words, the $1/\sigma_{L}$ expansion and the limit $r\to 0$ is not commutable.  Oscillatory behavior at small $r$ and small $\sigma_{L}$ may be the signal of that non-commutablity.  From quantitative point of view, most convenient value of the Wilson parameter would be a medium one around $r\sim 1/2$.

\section{Conclusion}
We first introduced an alternative summation scheme called square scheme where the expansion in $\delta$ is carried out to $\delta^{N}$ for all powers of $M^{-1}$.  Like the equalizer, the square scheme enhances the contribution of $M^{-k}$ by the factor $\Big(
\begin{array}{c}
N+k \\
k
\end{array}
\Big)$.    
Then, 
we have shown that the delta expansion in square scheme leads to Borel transform in the limit $M, N\to \infty$ with $M/N$ kept fixed.   In the large ${\cal N}$ limit of ${\cal N}$ component anharmonic oscillator, this limit was shown to be natural by explicit application of square schemes. 

In anharmonic oscillators and non-linear $\sigma$ model, conventional and square schemes can be useful for the simulation of the continuum limit via $1/M$ expansion.  In fact the accuracy of approximation obtained in BT limit is slightly worse than that obtained in conventional scheme employed in \cite{yam,hhy}.  However, it is important that in both truncation schemes, the double series in $M^{-1}$ and $\delta$ have been found to converge to the unique limit, the correct scaling, even in the $\delta\to 1$ limit.  In addition, flexibility and stability of the delta expansion under various truncation schemes was explicitly demonstrated.   For example, from the connection to Borel transform, we can make rough estimation in what cases delta expansion is effective or not.  One might guess that the expansion in $\delta$ in large $M$ series would not reflect the dilation around $M=0$, if there exists a singularity on positive real axis of $M$.  This is supported by the criterion of the Borel non-summability.   Fortunately, lattice models allowing the continuum limit undergo second order phase transition and the transition occurs in the $M\to 0$ limit and does not at $M>0$.   Hence, large $M$ expansion belongs to Borel summable type and the delta expansion in various truncation schemes would help us to capture the continuum scaling in other complex lattice models.  Thus, the mass in momentum space is a suitable parameter to express physical quantities. 

$2d$ Gross-Neveu model is an example where the conventional truncation does not work and the square truncation, especially its BT limit, is necessary.  This is because the linear divergence due to the Wilson term must be cancelled by the perturbative counter term while the original quantum correction needs to be expanded in $1/\sigma_{L}$.   Using Borel transform, we found that the value of Wilson parameter around $1/2$ is suitable for  the quantitative study of the scaling and the dynamical mass evaluation.
%
%

\end{document}